\documentclass[a4paper]{jpconf}
\usepackage{graphicx}
\begin{document}
\title{The neutrino masses and the change of allowed parameter region in universal extra dimension models}

\author{Shigeki Matsumoto$^a$, Joe Sato$^b$, Masato Senami$^c$, and Masato Yamanaka$^b$}

\address{ $^a${\it Institute for International Advanced Interdisciplinary Research,
     
Tohoku University, Sendai, Miyagi 980-8578, Japan

$^b${\it Department of Physics, Saitama University, 
     
Shimo-okubo, Sakura-ku, Saitama, 338-8570, Japan

}

$^c${\it ICRR, University of Tokyo, Kashiwa, Chiba 277-8582, Japan

}
}}

\ead{masa@krishna.phy.saitama-u.ac.jp}

\begin{abstract}
 Relic abundance of dark matter is investigated in the framework of universal extra dimension 
models with right-handed neutrinos. These models are free from the serious Kaluza-Klein (KK) 
graviton problem that the original universal extra dimension model possesses. The first KK 
particle of the right-handed neutrino is a candidate for dark matter in this framework. When 
ordinary neutrino masses are large enough such as the degenerate mass spectrum case, the dark 
matter relic abundance can change significantly. The scale of the extra dimension consistent 
with cosmological observations can be 500 GeV in the minimal setup of universal extra dimension 
models with right-handed neutrinos. 
\end{abstract}

\section{Introduction and UED models} 

As is well known, there is dark matter in this universe, and many models beyond the Standard Model (SM) 
have been proposed to explain the dark matter. Among those, Universal Extra Dimension (UED) models 
are one of interesting candidates for new physics. However, because only a short term passes since UED 
model was proposed, many subject remains which should be investigated. We have solved problems inherent 
in these models, and calculated the allowed parameter for UED models by estimating the dark matter relic 
abundance. This proceeding is based on our works \cite{Matsumoto:2007dp} and \cite{Matsumoto:2006bf}.

First, we briefly review UED model. This model is the simplest extension of the SM in the five-dimensional 
space-time, where the extra dimension is compactified on an $S^1/Z_2$ orbifold with the radius $R$. By 
expanding SM fields on fifth-dimensional space time, we can show that there are infinite tower of excited 
states. These excited states are called Kaluza-Klein (KK) particles. The SM particles and their KK 
particles have identical gauge charges. All interactions relevant to KK particles are determined by the 
SM Lagrangian. 

Though the original UED model is phenomenologically good model, the model has two shortcomings. The 
first one is called the KK graviton problem. In the parameter region where (extra dimension scale 
$1/R$ ) $<$ 800 GeV, the KK graviton is the lightest KK particle (LKP), and the next LKP (NLKP) is 
the KK photon. Hence, the KK photon decays into a KK graviton and a photon at very late time due to 
gravitational interactions. This fact leads to the serious problem; KK photons produced in the early 
universe decay into photons in the late universe, and these photons distort the CMB spectrum or the 
diffuse photon spectrum.

The second problem is the absence of neutrino masses. Since UED model has been constructed as minimal 
extension of the SM, neutrinos are treated as massless particles. However we know that it is not true. 
Therefore we must introduce neutrino masses into UED models.

\section{Solving the problems} 
 
In order to solve these problems, we introduce the right-handed neutrinos into UED models, and assume
that they form Dirac mass with ordinary neutrinos. The neutrino masses are expressed as 
\begin{eqnarray}
 {\cal L}_{\nu{\rm -mass}}
 = y_\nu \bar N L \Phi
 + {\rm h.c.}~,
 \label{Neutrino Yukawa}
\end{eqnarray}
where $N$ is the right-handed neutrino, $L$ is the left-handed (doublet) lepton. Thus the second problem,
the absence of the neutrino masses, are clearly solved. Once we introduce right-handed neutrinos in the 
MUED model, their KK particles automatically appear in the spectrum. The mass of the first KK particle 
of the right-handed neutrino, $N^{(1)}$, is estimated as
\begin{eqnarray}
 m_{N^{(1)}}  \simeq \frac{1}{R} + O \left( \frac{m_{\nu }^2}{1/R} \right)~.
\end{eqnarray}
By comparing the mass of $N^{(1)}$ with the mass of other KK particles, we can see that $N^{(1)}$ is 
the NLKP, and the KK photon is the next to next lightest KK particle.

The existence of the $N^{(1)}$ NLKP changes the late time decay of the KK photon. In the models with the 
KK right-handed neutrino, the KK photon dominantly decays into $N^{(1)}$ and SM left-handed neutrino at 
tree level. Calculating the decay rate of KK photon decay modes, we can show that the decay of the KK 
photon is governed by the process, $\gamma^{(1)}\rightarrow N^{(1)}\bar{\nu}$. On the other hand, the 
dominant decay mode associated with a photon, $\gamma^{(1)}\rightarrow G^{(1)}\gamma$, comes from the 
Planck suppressed process, and its branching ratio is
\begin{eqnarray}
 {\rm Br_{X\gamma}}
 =
 \frac{\Gamma(\gamma^{(1)} \rightarrow G^{(1)}\gamma) }{\Gamma(\gamma^{(1)} \rightarrow N^{(1)} \bar \nu)}
 =
 5 \times 10^{-7}
 \left(\frac{1/R}{500 {\rm GeV}}\right)^3
 \left(\frac{0.1 {\rm eV}}{m_\nu}\right)^2
 \left(\frac{\delta m}{1 {\rm GeV}}\right) ~.
\label{eq:br}
\end{eqnarray}
As a result, by introducing the right-handed neutrino into UED models, neutrino masses are introduced, and
problematic high energy photon emission is highly suppressed. Therefore, two problems in UED models have 
been solved simultaneously.

\section{$N^{(1)}$ dark matter and parameter determination } 

$N^{(1)}$ cannot decay because it is forbidden by the kinematics. Since $N^{(1)}$ is neutral,massive, 
and stable, $N^{(1)}$ can be dark matter candidate. Hence, if we introduce the right-handed neutrinos 
into UED models, dark matter changes from the LKP KK graviton to the NLKP KK right-handed neutrino.
The KK graviton dark matter is produced from only decoupled KK photon decay, while $N^{(1)}$ dark matter
is produced from decoupled KK photon decay and from the high temperature thermal bath. As a result, in our 
model, there are additional contribution to the dark matter relic abundance. As total dark matter number 
density becomes large, the dark matter mass, $\sim 1/R$, becomes small. Therefore, in order to determine 
the compactification scale $1/R$, we must reevaluate the number density of $N^{(1)}$ dark matter.

Because the number density of the decoupled KK photon has been calculated in previous works 
\cite{Kakizaki:2006dz}, we can see the $N^{(1)}$ number density produced from the decoupled KK photon 
decay. So, we need to calculate the 
$N^{(1)}$ number density produced from the thermal bath. To do this, we must take account of the thermal 
effects. If the thermal bath temperature is high enough, the mass of a particle receives a correction 
which is proportional to square of thermal bath temperature. For example, the thermal mass of the KK Higgs 
boson $\Phi^{(n)}$ is given by the following formula, 
\begin{eqnarray}
 m_{\Phi^{(n)}}^2(T)
 =
 m_{\Phi^{(n)}}^2(T=0)
 +
 \left[
  a (T) \cdot 3\lambda_h
  +
  x (T) \cdot 3 y_t^2
 \right] \frac{T^2}{12}~.
 \label{eq:thermalmass}
\end{eqnarray}
Coefficients $a(T)$ and $x(T)$ are determined by evaluating how many KK modes contribute to the correction 
at the temperature $T$. Taking account of such a thermal masses, we calculated the $N^{(1)}$ number density.

\begin{figure} [t]
\begin{center}
 \scalebox{.58}{\includegraphics*{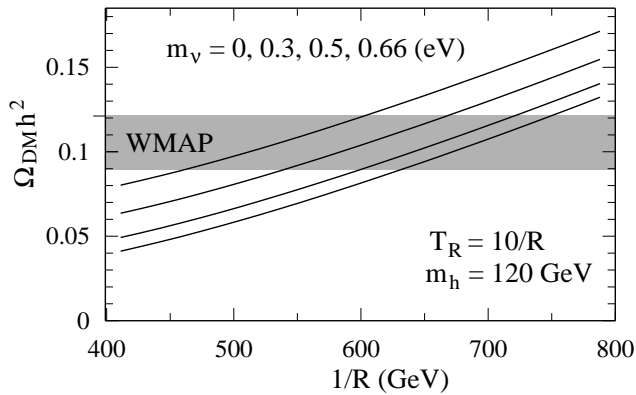}}
 \caption{The dependence of the abundance on $m_\nu$ with fixed reheating 
          temperature $ T_R = 10 /R $ and $ m_h = 120 $ GeV. The solid lines
          are the relic abundance for $ m_\nu = 0, 0.3, 0.5, 0.66 $ eV from 
          bottom to top.  The gray band represents the allowed region from 
          the WMAP observation at the $2 \sigma$ level}
 \label{fig:mnu}
\end{center}
\end{figure}

Finally, we show the numerical result. In Fig.1, the neutrino mass dependence of the abundance. The 
horizontal axis is the compactification scale of the extra dimension, and the vertical axis is the relic
abundance of the dark matter. The solid lines correspond to the result with $ m_\nu = 0, 0.3, 0.5, 0.66 $
eV from bottom to top. As shown in this figure, as neutrino mass becomes large, compactification scale can 
be less than 500 GeV. If compactification scale is less than 500 GeV, in ILC experiment, n=2 KK particles 
can be produced. Such a particles are very important for discriminating UED from SUSY at collider experiment.

\section{ Summary } 

We have solved two problems in UED models (absence of the neutrino mass, energetic photon emission) 
by introducing the right-handed neutrinos. We have shown that by introducing right-handed neutrino, 
the dark matter is the KK right-handed neutrino, and we have calculated the relic abundance of the 
KK right-handed neutrino dark matter. In the UED model with right-handed neutrinos, the compactification 
scale $1/R$ can be less than 500 GeV. This fact has important consequence on the 
collider physics, in particular on future linear colliders, because first KK particles can be produced
in a pair even if the center of mass energy is around 1 TeV.

\section*{References}



\begin{thebibliography}{99} 

\bibitem{Matsumoto:2007dp}
  S.~Matsumoto, J.~Sato, M.~Senami and M.~Yamanaka,
  {\it Phys. Rev.}\  D {\bf 76}, 043528 (2007)


\bibitem{Matsumoto:2006bf}
  S.~Matsumoto, J.~Sato, M.~Senami and M.~Yamanaka,
  {\it Phys. Lett.}\  B {\bf 647}, 466 (2007)
  

\bibitem{Kakizaki:2006dz}
  M.~Kakizaki, S.~Matsumoto and M.~Senami,
  {\it Phys. Rev.}\  D {\bf 74} (2006) 023504; 
  
  S.~Matsumoto and M.~Senami,
  {\it Phys. Lett.}\  B {\bf 633} (2006) 671





\end{thebibliography}
\end{document}